\documentclass[twocolumn, showpacs]{revtex4}

\usepackage{graphicx}
\usepackage{dcolumn}
\usepackage{bm}

\begin{document}


\title{Enhanced Elasticity and Soft Glassy Rheology of a Smectic in a Random 
Porous Environment}

\author{Ranjini Bandyopadhyay$^{1}$, Dennis Liang$^{1}$, Ralph H. 
Colby$^{2}$, James L. Harden$^{3}$ and Robert L. Leheny$^{1}$}

\affiliation{$^{1}$Department of Physics and Astronomy, Johns Hopkins 
University, Baltimore, MD 21218, USA\\
$^{2}$Department of Materials 
Science and Engineering, 
Pennsylvania State University, University Park, 
PA 16802, USA\\
$^{3}$Department of Chemical and Biomolecular 
Engineering, Johns Hopkins 
University, Baltimore, MD 21218, 
USA\\}

\date{\today}

\begin{abstract}
We report studies of the frequency dependent shear modulus, 
$G^*(\omega)=G'(\omega)+iG''(\omega)$, of the liquid crystal octylcyanobiphenyl 
(8CB) confined in a colloidal aerosil gel.  With the onset of smectic 
order, $G'$ grows approximately linearly with decreasing temperature, 
reaching values that exceed by more than three orders of magnitude the 
values for pure 8CB.  The modulus at low temperatures possesses a 
power-law component, $G^*(\omega) \sim  \omega^\alpha$, with exponent $\alpha$ 
that approaches zero with increasing gel density.  The amplitude of 
$G'$ and its variation with temperature and gel density indicate that the 
low temperature response is dominated by a dense population of defects 
in the smectic.  In contrast, when the 8CB is isotropic or nematic, the 
modulus is controlled by the elastic behavior of the colloidal gel.  

\end{abstract}


\maketitle

Impurities and other forms of quenched disorder can profoundly affect 
the mechanical properties of materials.  For instance, introducing 
nanometer-scale inclusions into polycrystalline solids is a well-developed 
approach for enhancing strength by pinning defects whose mobility 
dictates mechanical compliance~\cite{mcclintock}.  Many complex fluid phases 
can similarly possess topological defects that influence their 
viscoelasticity.  Important examples are lamellar systems -- such as smectic 
liquid crystals, diblock copolymers, and surfactant mesophases -- that 
are characterized by one-dimensional ordering from the stacking of 
liquid-like planes.  Unlike the case of crystals where impurities can be 
frozen into the lattice, introducing quenched disorder into such fluids is 
a technical challenge.  In some cases, annealed disorder in the form of 
colloids suspended in lamellar systems can anchor line defects into a 
crosslinked network that enhances the elastic 
response~\cite{ramos,basappa}.  

One systematic approach for introducing quenched disorder to a smectic 
is to confine the fluid in a random porous 
medium~\cite{clark-science,germano,finotello-sil,paperI,paperII,clegg8OCB,dennis,leo}.  Study of 
such systems has largely concerned the influence of the environment on 
the phase behavior and smectic ordering.  Theory and experiment have 
concluded that the smectic phase is highly sensitive to the quenched 
disorder.  An interesting example is a smectic confined by an aerosil gel, 
which is a compliant hydrogen-bonded gel formed from nanometer-scale 
silica 
colloids~\cite{germano,finotello-sil,paperI,paperII,clegg8OCB,dennis}.  Consistent with theory~\cite{leo,olmsted}, x-ray scattering on 
liquid crystal - aerosil composites has revealed that the gel introduces 
random field effects that 
replace the thermodynamically sharp nematic to smectic transition with 
the growth of short-range smectic 
correlations~\cite{paperI,clegg8OCB,dennis}.  Theory further predicts that quenched disorder impacts the 
mechanical behavior of smectics by introducing enhanced anomalous 
elasticity~\cite{leo}.  We present rheometry studies  that characterize how the 
viscoelastic properties of a smectic are influenced by such disorder.  
Using weakly bonded aerosil gels enables us to isolate the mechanical 
response of the smectic.  We find that the quenched disorder greatly 
increases the shear modulus over that of the pure smectic and leads to 
weak power-law behavior characteristic of soft glassy rheology.  

Samples were prepared by forming aerosil gels within the liquid crystal 
octylcyanobiphenyl (8CB) following established 
procedures~\cite{germano}.  The 8CB (Frinton Labs) had a quoted purity of 99.4\%.  Pure 8CB 
undergoes an isotropic to nematic transition at $T_{NI}$ = 313.98 K and a 
nematic to smectic-A transition in which the molecules form a density 
wave with period $d_0 = 3.4$ nm at $T_{NA}$ = 306.97 K~\cite{germano}.  
The aerosil (DeGussa Corp., type 300) consists of 7 nm SiO$_2$ 
particles that are hydrophilic due to a high density of surface hydroxyl 
groups.   Degassed 8CB and dried aerosil were mixed with high purity acetone 
and sonicated for several hours to achieve a uniform dispersion.  The 
mixtures were then heated to 320 K to evaporate the acetone slowly.  
During this process the aerosil aggregated via hydrogen bonding.  The 
resulting composites were uniform, thixotropic solids comprised of 8CB 
permeated by a colloidal gel.  Gels formed in this way have a fractal 
dimension near 2.1~\cite{germano_SAXS} and a broad distribution of void 
sizes with a mean pore chord $l_0$ that varies with aerosil concentration 
as $l_0 = 6.7/\rho_S$ nm, where $\rho_S$ is the aerosil density in grams 
SiO$_2$/cm$^3$ 8CB~\cite{paperII}.  Samples were prepared with $\rho_S 
= $ 0.035  to $\rho_S =$ 0.10, corresponding to $l_0 \approx 56d_0$ to 
$l_0  \approx 19d_0$ and aerosil volume fractions of 0.016 to 0.043.  
For $\rho_S >$ 0.10 the composites change from a pasty to a brittle 
consistency that makes rheometry measurements difficult.  

The gel affects both the liquid crystal's nematic and smectic behavior.  
High porosity gels weakly perturb the orientational ordering, as 
evidenced by nematic correlation lengths that exceed $l_0$ by a factor of 30 
or more for $\rho_S < $ 0.10 ~\cite{bellini_static}.  The observed 
random field behavior of the smectic~\cite{paperI} is understood by 
considering how the randomly positioned gel strands impose steric hinderance 
on the liquid crystal molecules, effectively forcing the phase of the 
smectic mass density wave to different values at different points in 
space.  The sensitivity of the smectic ordering to the gel is reflected in 
the smectic correlation lengths, which are nearly an order of magnitude 
smaller than the nematic correlation lengths~\cite{paperI}.

The frequency-dependent shear modulus, 
$G^*(\omega)=G'(\omega)+iG''(\omega)$, of the 8CB-aerosil composites was determined using a stress 
controlled rheometer (Paar Physica MCR300) in a cone and plate geometry 
equipped with a Peltier device that maintained temperature stability to 
within $\pm$ 0.01 K.  Temperature gradients across the sample were 
estimated at less than 0.2 K.  
Figure 1(a) shows the storage modulus $G'(\omega= 1$ rad/s) as a 
function of strain amplitude $\gamma_0$ for  $\rho_S =$ 0.10 at temperatures 
above and below the smectic transition.  When the 8CB possesses smectic 
order, the response depends weakly on $\gamma_0$ to the smallest 
strains.  However, measurements of $G^*(\omega)$ with different $\gamma_0$ 
showed only this slight variation in magnitude; all salient features, 
particularly the dependence on $\omega$ and temperature, were independent 
of $\gamma_0$ for $\gamma_0 \leq 0.007$.  The results presented below 
were obtained with  $\gamma_0 = 0.003$.  

\begin{figure}
\includegraphics[scale=0.65]{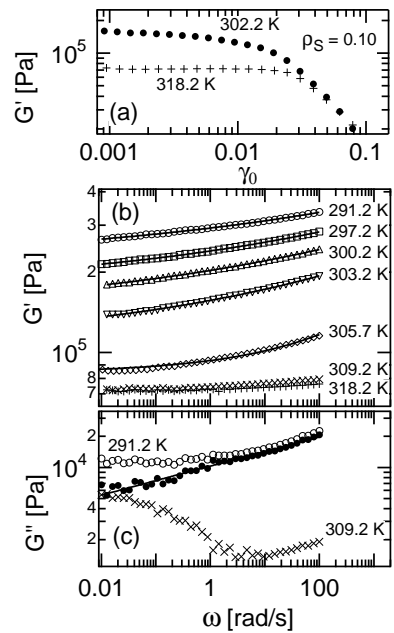}
\caption{ (a) Storage modulus $G'$ of 8CB confined in an aerosil gel 
with $\rho_{S}=0.10$ g/cm$^{3}$ at $\omega = 1$ rad/s as a function of 
strain amplitude $\gamma_0$ below (solid circles) and above (+) the 
smectic transition temperature, $T_{NA}$.  (b) $G'(\omega)$ at several 
temperatures for $\gamma_0 = 0.003$.  Lines are results of fits to Eq.~(1).  
(c) Loss modulus $G''(\omega)$ below (open circles) and above (Xs) 
$T_{NA}$.  The solid circles display the difference between the two 
spectra, and the line is the result of a power-law fit.}
\label{fig1}
\end{figure}

Figures 1(b) and 1(c) show $G'(\omega)$ and the loss modulus 
$G''(\omega)$ respectively at several temperatures for $\rho_S = 0.10$.  Above 
307 K the modulus obeys trends of a weak elastic solid, with $G'(\omega)$ 
constant and $G'(\omega) >> G''(\omega)$, and displays little 
temperature dependence.  In particular, the isotropic to nematic transition near 
314 K has no measurable effect on $G^*(\omega)$.  We associate this 
high temperature mechanical response with that of the aerosil gel in a low 
viscosity liquid crystal solvent.  Measurements of aerosil gels in the 
simple organic liquid dibutyl phthalate give quantitatively similar 
$G^*(\omega)$, supporting this interpretation.  The storage modulus 
$G_{g}$ above the smectic transition temperature varies with aerosil density 
roughly as $G_{g} \sim \rho_S^{4.8 \pm 0.8}$, consistent with behavior 
expected for a colloidal gel with fractal dimension near 
2.1~\cite{buscall}.  With the onset of smectic ordering in the 8CB at low 
temperatures, $G'(\omega)$ increases dramatically and obtains a clear frequency 
dependence.  We attribute the growth in $G^*(\omega)$ to contributions 
from smectic defects.  The solid lines in Fig.~1(b) are the results of 
fits to a power-law contribution plus a frequency-independent background
\begin{equation}
G'(\omega) = G_g+G_s+\beta_{s}\omega^{\alpha}.
\end{equation}
\noindent
where $G_g$ is the contribution from the gel, assumed independent of 
$\omega$ and temperature, and $G_s(T)$ is the frequency-independent 
contribution from the smectic phase.  This form describes accurately the 
storage modulus at all densities of aerosil and at all temperatures.  

The inset to Fig.~2 displays the amplitudes for the power-law and 
frequency-independent contributions, $\beta_s$ and $G_s$, as a function of 
temperature for $\rho_S$ = 0.10.  These parameters show qualitatively 
the same temperature dependence for all $\rho_S$ except that the ratio 
$\beta_s/G_s$ increases with increasing $\rho_S$.  The exponent $\alpha$ 
characterizing the power-law dependence displays no systematic 
temperature dependence below the pseudocritical point marking the onset of 
static short-range smectic order~\cite{paperI,paperII} but does vary 
systematically with $\rho_S$.  Figure 2 displays the value of $\alpha$ 
averaged over temperature as a function of $\rho_S$ wherein $\alpha$ 
decreases from $\alpha \approx 0.25$ at $\rho_S = 0.035$ to $\alpha \approx 
0.1$ at $\rho_S = 0.10$.   Such weak power-law behavior is characteristic 
of a variety of complex fluids including foams, emulsions, particulate 
suspensions and slurries.  In a recent model  Sollich and 
coworkers~\cite{sollich} provide a unifying theoretical framework for this ``soft 
glassy rheology'', arguing that the response is a general consequence of 
structural disorder and metastability.  In this model $\alpha$ serves 
as an effective noise temperature, with systems entering a glass 
transition as $\alpha \rightarrow 0$.   (The exponent $x$ in  
Ref.~\cite{sollich} is $x= \alpha+1$.)  Thus, the variation of $\alpha$ with $\rho_S$ 
points to increasingly glassy dynamics in the smectic with increased 
quenched disorder.  This notion of glassy behavior further connects the 
rheology with the slow dynamics observed with dynamic light scattering 
for 8CB in aerosil~\cite{mertelj}.  However, we note the simultaneous 
growth of the frequency-independent component $G_s$ implies some portion 
of the 8CB is frozen for all $\rho_S$.  

\begin{figure}
\includegraphics[scale=0.65]{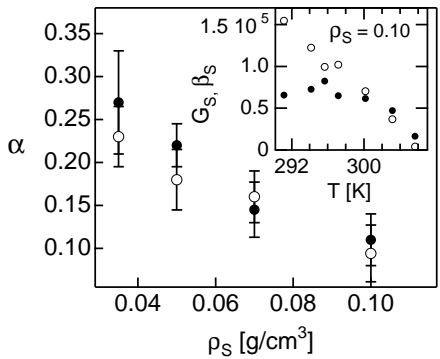}
\caption{The 
exponent $\alpha$ characterizing the power-law contribution to 
$G'(\omega)$ (solid circles) and $G''(\omega)$ (open circles) for smectic 8CB 
confined by an aerosil gel as a function of gel density.  The inset 
displays results for $G_s$ (open circles) and $\beta_s$ (filled circles) 
corresponding to the amplitudes of the frequency-independent and 
power-law contributions to $G'(\omega)$, respectively, for 8CB in a gel with 
$\rho_{S}=0.10$ g/cm$^{3}$ at temperatures below the onset of smectic 
order in the 8CB.}
\label{fig2}
\end{figure}

As expected for soft glassy rheology~\cite{sollich}, the weak power-law 
behavior also appears in the loss modulus.   Figure 1(c) displays 
$G''(\omega)$ at low temperature in the smectic phase of 8CB and at high 
temperature in the nematic phase, where contributions from the gel 
presumably dominate.  With the high temperature modulus subtracted from the 
low temperature spectrum, the remaining smectic contribution, shown by 
the solid symbols in Fig.~1(c), displays power-law behavior.  As shown in 
Fig.~2,  fits to such results, $G''(\omega) \sim G'(\omega) \sim 
\omega^{\alpha}$, provide values for $\alpha$ that agree closely with those 
obtained from $G'(\omega)$.

The amplitude and frequency dependence of $G'(\omega)$ for smectic 8CB 
in aerosil differ from those of other unaligned lamellar complex 
fluids.  A wide range of systems including lyotropic and thermotropic 
smectics (without quenched disorder), surfactant-based phases, and diblock 
copolymers exhibit a frequency response with two contributions:  
$G'(\omega) = G_0+\beta_d\omega^{1/2}$.  The plateau modulus $G_0$ arises from 
the elasticity of static defects while the frequency-dependent term 
results from the bulk response of the material.  Specifically, 
$\beta_d\omega^{1/2}$ with $\beta_d=(\pi/24\sqrt{2})\sqrt{(B\eta)}$, where $B$ is 
the compression modulus and $\eta$ is an effective viscosity, describes 
the response of lamellar regions with the layer normal oriented so that 
strain necessitates layer compression~\cite{kawasaki}.  While the 
relative strength of these terms varies with material parameters and defect 
structures, most studies on unaligned lamellar systems capture this 
characteristic response in their experimental frequency range.   In 
particular, experiments on pure unaligned 8CB display a plateau modulus $G_0 
\approx$ 100 Pa and a $\omega^{1/2}$ contribution with $\beta_d \approx 
30$ Pa$\cdot$s$^{1/2}$~\cite{larson,colby}, consistent with the room 
temperature values $B \approx 2\cdot10^6$ Pa and $\eta \approx$ 0.1 
Pa$\cdot$s.  The response we observe for 8CB in aerosil  is {\it more than 
three orders of magnitude larger} than $G'(\omega)$ of pure 8CB.  The 
greatly enhanced modulus in the presence of disorder implies that 
contributions from the bulk fluid are negligible.  Instead, we infer that the 
shear response derives from a dense population of defects that form in 
the smectic with the destruction of smectic correlations by the 
disorder.  Constraints imposed on defect motion by the confining gel provide 
rigidity that dominates the glassy response to shear.

Evidence for soft glassy rheology and the approach to a glass 
transition with increasing $\rho_S$ comes not only from the weak power-law 
behavior of the modulus but also from the ratio $G''/G'$.  Figures 3(a) and 
3(b) show the temperature dependence of $G'$ and $G''$, respectively, 
at $\omega=1$ rad/s for varying $\rho_S$.  Within the soft glassy 
rheology model $G''/G' \sim  \alpha$~\cite{sollich}.  Figure 3(c) shows 
$G''/G'$ scaled by $\alpha$.  At temperatures below the pseudocritical 
region, the ratios collapse onto a single scaling function.  This scaling 
demonstrates how the single parameter $\alpha$ accounts for the 
variations in mechanical response with changing $\rho_S$.  Thus, a direct 
correspondence exists between the strength of random fields in the system, as 
parameterized by $\rho_S$~\cite{paperI,paperII}, and the exponent 
$\alpha$ that is central to soft glassy rheology.  

\begin{figure}
\includegraphics[scale=0.65]{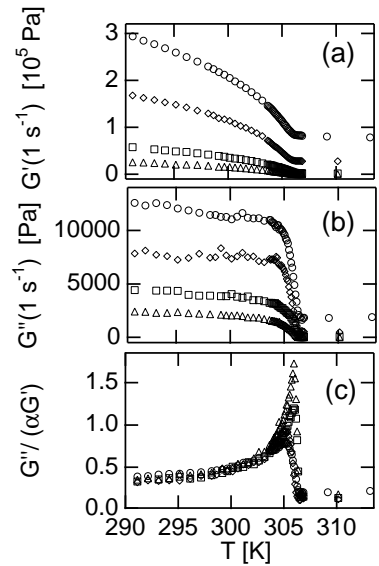}
\caption{(a) Storage modulus $G'(T)$ and (b) loss modulus $G''(T)$ at 
$\omega = 1$ rad/s for 8CB in an aerosil gel with $\rho_{S}=0.035$ 
g/cm$^{3}$ (triangles), 0.05 g/cm$^{3}$ (squares), 0.07 g/cm$^{3}$ 
(diamonds), and 0.10 g/cm$^{3}$ (circles).  (c) The ratio $G''/G'$ scaled by the 
exponent $\alpha$ for the four densities.}
\label{fig3}
\end{figure}

The variation of $G'$ with $\rho_S$ indicates that smectic line defects 
dominate this soft glassy rheology.  In analogy with rubber elasticity, 
the elastic response of a static defect network should vary as $G_s 
\approx \tau/d^2$ where $\tau$ is the defect line tension and $d$ is a 
typical spacing between defects~\cite{ramos}.  A plausible scale for $d$ 
is set by the porosity of the gel.  Assuming $d \approx \xi \sim l_0$, 
where $\xi$ is the smectic correlation length and $l_0$ is the gel's 
mean pore chord~\cite{paperI,germano}, we expect $G_s \sim l_0^{-2}  \sim 
\rho_S^2$.  The inset to Fig.~4 displays $G_s$ at 292.6 K, well below 
the nematic to smectic transition, as a function of $\rho_S$.  The solid 
line in the inset represents the result of the best fit to a power law,  
$G_s \sim \rho_S^{1.9\pm0.2}$, consistent with this expectation.  

\begin{figure}
\includegraphics[scale=0.65]{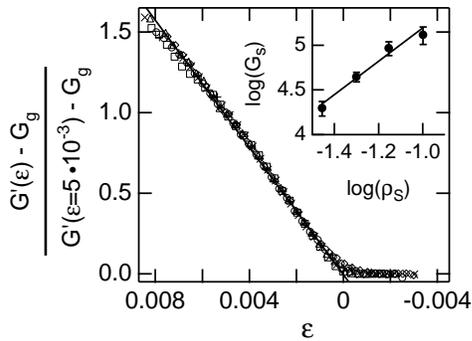}
\caption{Smectic contribution to $G'$ near the pseudotransition 
temperature that marks the onset of smectic order normalized by its value at 
the reduced temperature $\epsilon=5\cdot10^{-3}$ for 8CB in an aerosil 
gel.  Included are $G'(\omega = 1)$ for $\rho_{S}=0.035$ g/cm$^{3}$ 
(triangles), 0.05 g/cm$^{3}$ (squares), 0.07 g/cm$^{3}$ (diamonds), and 
0.10 g/cm$^{3}$ (circles) as well as $G'(\omega = 0.1)$ (crosses) and 
$G'(\omega = 40)$ (X's) for 0.07 g/cm$^{3}$.  The solid line represents a 
linear dependence on $\epsilon$.  The inset displays the values of $G_s$ 
at 292.6 K with the solid line showing the fit result $G_s \sim 
\rho_S^{1.9\pm0.2}$.}
\label{fig4}
\end{figure}

The temperature dependence of $G'$ similarly indicates defects dominate 
the mechanical response.  
Figure 4 shows the smectic contribution to $G'$ as a function of 
reduced temperature, $\epsilon\equiv(T_{NA}-T)/T_{NA}$, normalized by its 
value at $\epsilon=5\cdot10^{-3}$~\cite{reduced_T}.  
Included in the plot are data from Fig.~3(a) covering $G'(\omega = 
1,\epsilon)$ for four aerosil densities as well as $G'(\omega = 
0.1,\epsilon)$ and $G'(\omega = 40,\epsilon)$ for $\rho_{S}=0.07$.  The data 
collapse onto the solid line which displays the relation $(G'(\epsilon) - 
G_g) \sim \epsilon^z$ with $z=1$.  This scaling demonstrates that the 
smectic contribution to the mechanical response is associated with 
critical phenomena of the nematic to smectic transition.  As mentioned above, 
analogy with rubber elasticity predicts $G' \sim \tau$, where $\tau$ is 
the defect line tension.  For screw dislocations, which are prominent 
in smectics~\cite{screws}, $\tau = Bb^4/128\pi^3r_c^2$, where $b=md_0$ 
is the Burger's vector of integer strength $m$ and $r_c$ is the defect 
core radius~\cite{kleman}.  For 8CB near $T_{NA}$, $B \sim 
\epsilon^{0.4}$~\cite{marcerou2} and $r_c^{-2} \sim \psi^2 \sim \epsilon^{0.5}$, 
where $\psi$ is the smectic order parameter~\cite{paperI}.  Thus,  $\tau 
\sim \epsilon^{0.9}$ in close agreement with $z \approx 1$ we observe 
for $G'$, suggesting that screw dislocations play a predominant role in 
dictating the shear response.  A more quantitative assessment of the 
prominence of screw dislocations depends on the typical Burger's vector 
strength.  For a moderate value, $m=4$, along with the low temperature 
values $B \approx 10^7$ Pa~\cite{marcerou2}, $r_c \approx 1$ nm, and 
$G_s$ from the inset of Fig.~4, the relation $G_s \approx \tau/d^2$ leads 
to a plausible defect spacing $d \approx 0.5 l_0$, independent of 
aerosil density.


Regardless of the precise connection between the temperature dependence 
of $G'$ and quenched disorder, the observed enhanced elasticity and 
soft glassy rheology of 8CB confined in aerosil provides a new perspective 
on effects of quenched disorder in complex fluids.  Unlike for many 
soft glassy systems, the defects that drive this rheology provide a clear 
physical origin for the disorder and metastability in the smectic.  
Furthermore, the quenched disorder that the gel imposes on the smectic has 
been convincingly characterized in terms of random 
fields~\cite{paperI,leo}.  Thus, this mechanical behavior connects a classic model of 
quenched disorder, from random fields, with a new paradigm of glassy 
behavior in complex fluids.

We thank B. Erwin and R. Patil for assistance and Z. Te\v{s}anovi\'{c} 
for helpful discussions.  Funding was provided by the NSF 
(DMR-0134377).

\newpage


\begin{thebibliography}{100}
\bibitem{reduced_T}
The values for $T_{NA}$ 
were 306.16 K for $\rho_S = 0.035$, 306.35 K for $\rho_S = 0.05$, 
305.93 K for $\rho_S = 0.07$, and 306.18 K for $\rho_S = 0.10$, consistent 
with  previous studies [5,7]
\end{thebibliography}
\end{document}